\documentstyle[11pt,epsfig]{report}
\def\thebibliography#1{\leftline{\Large\it References}\list
  {[\arabic{enumi}]}{\settowidth\labelwidth{[#1]}\leftmargin\labelwidth
    \advance\leftmargin\labelsep
    \usecounter{enumi}}
    \def\newblock{\hskip .11em plus .33em minus .07em}
    \sloppy\clubpenalty4000\widowpenalty4000}

\newcommand{\be}{\begin{eqnarray}}
\newcommand{\ba}{\begin{array}}
\newcommand{\ea}{\end{array}}
\newcommand{\ee}{\end{eqnarray}}

\newcommand{\dslash}{\partial \hskip -0.5em /}

\newcommand{\vslash}{v \hskip -0.5em /}
\newcommand{\aslash}{a \hskip -0.5em /}

\newcommand{\nslash}{{\rm n} \hskip -0.5em /}
\newcommand{\bD}{{\bf D}}
\newcommand{\bDp}{{\bf D}^{(\pi)}}

\newcommand{\La}{{\cal L}}
\newcommand{\A}{{\cal A}}

\newcommand{\bjlim}{{\stackrel{\scriptstyle{\rm Bj}}
{\textstyle\longrightarrow}}}

\newcommand{\xipl}{\vec{\xi}\hskip-0.6mm
+\hskip-0.6mm\lambda\hat{e}_3}
\newcommand{\ximl}{\vec{\xi}\hskip-0.6mm
-\hskip-0.6mm\lambda\hat{e}_3}
\newcommand{\tauom}{\vec{\tau}\hskip-0.3mm
\cdot\hskip-0.3mm\vec{\Omega}}

\begin{document}
\begin{center}
{\Large\bf On the Structure Functions of Mesons and Baryons \\
in a Chiral Quark Model}\footnote{ Talk given at the Mini-Workshop
``Hadrons as Solitons'', Bled (Slovenia), July 9-17, 1999.}
\end{center}
\vskip 0.5cm
\centerline{ \Large
E. Ruiz Arriola \footnote{ e-mail: earriola@ugr.es }
}
\vskip 0.3cm
\begin{center} {\it
Departamento de F\'{\i}sica Moderna. 
Universidad de Granada. 
E--18071 Granada, Spain. } 
\end{center}
\date{\today}
\vskip 0.3cm
\centerline{\parbox[t]{15.8cm} 
{ \footnotesize {\bf ABSTRACT.}
This research summarizes work done by myself (Nucl.Phys.{\bf A641}
(1998)461), or in collaboration
with R. M. Davidson (Phys.Lett.{\bf B348}(1995)163) and H. Weigel
and L. Gamberg (Nucl.Phys.{\bf B560}(1999)xx). I will discuss
several topics related with the computation of structure functions in
the quark model in general and its perturbative evolution. In
particular, I address this topic in the Nambu--Jona--Lasinio model of
hadrons, where the nucleon is constructed as a soliton. I show
that the handling of the regularization procedure is crucial in order to
obtain exact scaling in the Bjorken limit and fulfillment of sum rules.
I also include some problems concerning the general validity of quark
model calculations.
}}
\vskip 0.15cm
\vfill
\noindent

\medskip

\bigskip
\stepcounter{chapter}
\leftline{\Large\it 1. Reminder on Deep Inelastic Scattering}
\medskip

Deep inelastic scattering (DIS) provides some
of the most convincing evidence for the quark sub--structure
\cite{Ro90} of hadrons. The main object of study which is
experimentally measured is the so-called hadronic tensor,
\be
W_{\mu \nu}^{ab}(p,q;s) = \frac{1}{4\pi}
\int d^4x e^{iq\cdot \xi}\,
\Big\langle p,s\Big| [ J_{\mu}^{a}(\xi),J_{\nu}^{b \dagger}(0)]
\Big| p,s \Big\rangle\, .
\label{hten1}
\ee
where $p,s$ are the hadron momentum and spin respectively,
$ J_{\mu}^{a}(\xi) $ is a hadronic vector or axial current
and $q$ is the momentum transfer. Introducing  the Lorentz invariants
$Q^2=-q^2$ and $x=Q^2/2p\cdot q$, one has on the basis of gauge and
relativistic invariance the following decomposition
(omitting parity violating contributions)
\be
W_{\mu \nu}^{ab}(p,q;s)
& = & \left(-g_{\mu \nu} + \frac{q_{\mu} q_{\nu}}{q^2}\right)
M_N W_{1} (x,Q^2)
\nonumber \\ & &
+\left(p_{\mu} - q_{\mu}\frac{p\cdot q}{q^2}\right)
\left(p_{\nu} - q_{\nu}\frac{p\cdot q}{q^2}\right)
\frac{1}{M_N} W_{2} (x,Q^2)
\nonumber \\ & &
+i\epsilon_{\mu \nu \lambda \sigma} \frac{q^{\lambda} M_N}{p\cdot q}
\left(\left[g_1(x,Q^2)+g_2(x,Q^2)\right]s^{\sigma}
-\frac{q\cdot s}{q\cdot p} p^{\sigma}
    g_2(x,Q^2) \right)\, .
\label{hten1a}
\ee
When a spin--zero hadron is considered, as {\it e.g.} the
pion, the polarized structure functions $g_1$ and $g_2$ are ignored.
Once the hadronic tensor is computed, the form factors are given
by suitable projections. Finally the leading twist contributions
to the structure functions are obtained from these form factors by
assuming the Bjorken limit:
\be
Q^2\to\infty
\qquad {\rm with}\qquad
x=Q^2/2p\cdot q \quad {\rm fixed}\, .
\label{bjl}
\ee
For the spin independent part the structure functions $f_i$ are the
linear combinations
\be
M_N W_1(x,Q^2)\, \bjlim\,  f_1(x)
\quad {\rm and} \quad
\frac{p\cdot q}{M_N}\,W_2(x,Q^2)\, \bjlim\,  f_2(x)= 2 x f_1(x)  \, .
\label{deff1f2}
\ee
where the last identity is a direct consequence of the spin 1/2 nature
of partons. The structure function is related to the quark and antiquark 
distribution functions, $q_i (x)$ and $\bar q_i(x)$ respectively by 
\be 
f_1(x) = {1\over 2} \sum_{i=u,d,s} e_i^2 \Big[ q_i (x) + \bar q_i (x) \Big]
\, , 
\ee
with $e_i$ the quark electromagnetic charges. 
Actually, exact scaling is only true up to logarithms,
$ {\rm log} (Q^2/\Lambda^2) $, due to perturbative QCD radiative
corrections. The most economic way of including them in a calculation is
by means of the DGLAP equations \cite{AP77}
which, using renormalization group
arguments, relate in a linear fashion the structure functions at a given
reference scale, say $Q_0^2$, to the scale of interest, $Q^2$,
\be
f_i (x,Q^2 ) = U(Q^2, Q^2_0 ) f_i (x,Q_0^2 ).
\ee
Here, $U(Q^2,Q_0^2)$ is a linear matrix operator, fulfilling the
properties $ U(Q_1^2 , Q_2^2 ) U(Q_2^2 , Q_3^2 ) = U(Q_1^2 , Q_3^2 ) $
and $ U(Q^2,Q^2)=1 $
In principle, the former
equation means that one only needs to know structure functions at a
given scale, since $U$ generates structure functions at all other
scales. So far $U$ can only be computed in perturbation theory.
Due to asymptotic freedom, this means
that in practice $U$ is reliable to relate high momentum scales,
although nobody really knows what high momentum means in the present
context; one is only left to practical convergence criteria between
one loop and two loop calculations. The form of scaling violations
predicted by QCD has been tested if one  relates experimentally
measured structure functions at different scales.

Expression (\ref{hten1}) is not always convenient for practical
calculations. In case the considered hadron represents the ground
state with the specific quantum numbers, the hadronic tensor will be
related to the forward virtual Compton amplitude
\be
W_{\mu \nu}^{ab}(p,q;s)=\frac{1}{2\pi}\, {\rm Im}\,
T_{\mu \nu}^{ab}(p,q;s)\, ,
\label{Comp1}
\ee
which is given as the matrix element of a time--ordered
product of the currents
\be
T_{\mu \nu}^{ab}(p,q;s) = i
\int d^4\xi e^{iq\cdot \xi}\,
\Big\langle p,s\Big| T\left\{J_{\mu}^{a}(\xi)
J_{\nu}^{b \dagger}(0)\right\}\Big| p,s \Big\rangle
\label{Comp2}
\ee
rather than their commutator. This time--ordered product has nicer
properties than the commutator, since it can be built by functional
differentiation with respect to external sources. Applying Cutkosky's
rules subsequently yields the absorptive part,
${\rm Im}\, T_{\mu \nu}$, which is to be
identified with the hadronic tensor (\ref{Comp1}).

In the parton model, an analysis of the virtual Compton forward
scattering amplitude leads, in the Bjorken limit,  to the so called
quark-target scattering formula \cite{JA85}. This formula is formally
correct, but it may require some regularization, or introduction of
form factors. It is obvious that not every choice of regularization or
form factor is consistent with the original Compton amplitude one
started with. This point is often disregarded in practical calculations.

The calculation of structure functions at a given scale remains a
challenge within QCD, and one may recourse to quark models to compute
them \cite{Ja75,model}. In all considered models one is forced to the
so-called valence quark approximation \cite{JR80}.
This hypothesis is based on the
observation that the momentum fraction carried by the valence quarks
increases for decreasing momentum scales. So, there is a momentum scale,
say $Q_0$, where the valence quarks carry all the momentum. In this
picture, one starts with only valence quarks and gluons and sea quark
distributions  are generated as radiative corrections by means of DGLAP
equations. Whether or not this assumption makes sense is analyzed below,
but one should say that it is routinely applied in practical
calculations.
There is another assumption which is usually made, related to the fact
that although models of hadrons deal with {\it constituent} quarks,
perturbative QCD evolution is applicable to {\it current} quarks. In my
opinion this point is at present not well understood, even in models
which exhibit dynamical symmetry breaking, i.e. which allow for both
kinds of quarks, and deserves further study.

\bigskip
\stepcounter{chapter}
\leftline{\Large\it 2. The NJL model in the Pauli-Villars
regularization}
\medskip

For practical calculations we adopt the
Nambu--Jona--Lasinio (NJL) model \cite{Na61} of quark flavor
dynamics. This model offers a microscopic and non-perturbative description
of dynamical chiral symmetry breaking for low energy non--perturbative
hadron physics, and is rather successful from a phenomenological point of
view. Mesons are described as quark-antiquark excitations of the vacuum
in terms of poles of the corresponding Bethe-Salpeter amplitude.
Baryons are described in the large $N_C $ limit of the model
as solitons. At a formal level the NJL quark distributions can be
identified with those of the parton model. However, a major obstacle is
that the bosonized NJL model contains quadratic divergences
which have to be carefully regularized, to comply with the symmetries of
the system. Actually, this point turns out to be crucial to ensure the
fulfillment of sum rules, i.e. weighted integrals of structure
functions in $x$ space which yield low energy matrix elements.

In the NJL chiral soliton model the issue of systematically regularizing
the nucleon structure functions is rather subtle and has been handled
in several ways. In a first approach the contributions from the
polarized vacuum  are neglected \cite{We96,We97}, since polarized vacuum
contributions to static nucleon properties turn out to be numerically
small, once the self-consistent soliton has been constructed. In other
approaches the quark-target scattering formula
has been regularized {\it a posteriori} \cite{Di96,Wa98}, in a treatment
which only works in the chiral limit since quadratic divergences
do not appear. This connection to static
properties for setting up the regularization description
does not provide a definite answer for structure functions
not having a sum rule; {\it i.e.}, structure functions whose
integral cannot be written as a matrix element of a local
operator.

The NJL  model is defined in terms
of quark fields by the Lagrangian~\cite{Na61}
\be
\La_{\rm NJL} = \bar q (i\dslash - m_0 ) q +
      2G_{\rm NJL} \left\{ (\bar q \frac{\vec{\tau}}{2} q )^2
      +(\bar q \frac{\vec{\tau}}{2} i\gamma _5 q )^2 \right\}
\label{NJL}
\ee
with the quark interaction described by a chirally symmetric
quartic potential. The current quark mass, $m_0$, parameterizes the
small explicit breaking of chiral symmetry. Using functional
techniques the quark fields can be integrated out in favor of
auxiliary mesonic fields, ${\cal M}=S+iP$. According to the
chirally symmetric interaction in the Lagrangian (\ref{NJL}),
$S$ and $P$ are scalar and pseudoscalar degrees of freedom,
respectively. This results in the bosonized action~\cite{Eb86}
\be
\A_{\rm NJL}=-iN_C {\rm Tr}_\Lambda\, {\rm log}\,
\left\{i\dslash - m_0 - \left(S + i\gamma_5 P\right)\right\}
+\frac{1}{4G}\int d^4x\,
{\rm tr}\left[{\cal M}{\cal M}^\dagger\right]\, .
\label{act0}
\ee
Here the `cut--off' $\Lambda$ indicates that the
quadratically and logarithmically
divergent quark loop requires regularization.
In order to compute properties of hadrons from the
action (\ref{act0}) a twofold procedure is in order. First,
formal expressions for the symmetry currents have
to be extracted. This is straightforwardly accomplished
by adding external sources with suitable quantum numbers
to the Dirac operator and
taking the appropriate functional derivatives. Ignoring effects
associated with the regularization, the currents would
be as simple as $\bar{q}\gamma_\mu (\gamma_5) t^a q$, with
$t^a$ being the appropriate flavor generator. Secondly,
hadron states are constructed from the effective
action which in turn allows one to calculate the relevant
matrix elements of the symmetry currents. For
the pion this will be a Bethe--Salpeter wave--function
which is obtained by expanding the action (\ref{act0})
appropriately in the fields $S$ and $P$. In the case
of the nucleon we will determine a soliton configuration
which after collective quantization carries nucleon quantum
numbers \cite{ANW}.

We employ the Pauli--Villars regularization scheme since it is
possible to formulate the bosonized NJL model
\cite{Na61} completely in Minkowski space. This is particularly
appropriate when applying Cutkosky's rules in order to
extract the hadronic tensor from the Compton amplitude. Also,
let us remind that in ref.\cite{Da95}, scaling for the pion structure
functions was accomplished in the Pauli-Villars regularization, and
not in the proper--time regularization. In this context, it has been
shown \cite{Da95bis} that in the Pauli-Villars regularization, unlike
the more customary proper--time regularization where cuts in the complex
plane appear \cite{Br96,Al96}, dispersion relations are fulfilled.
The Pauli--Villars regularization has been
considered before in this context both for mesons and solitons
and we refer to refs \cite{Ru91,Sc92,Da95} for more
details and results in this regularization scheme. We will
specifically follow ref \cite{Da95} because in that formulation
a consistent treatment solely in Minkowski space is possible.

The regularized action of the bosonized NJL model is then given
by\footnote{We denote traces of discrete
indices by ``${\rm tr}$'' while ``${\rm Tr}$'' also contains
the space--time integration.}
\be
\A_{\rm NJL}&=&\A_{\rm R}+\A_{\rm I}
+\frac{1}{4G}\int d^4x\,
{\rm tr}\left[S^2+P^2+2m_0S\right]
\label{act1} \\
\A_{\rm R}&=&-i\frac{N_C}{2}
\sum_{i=0}^2 c_i {\rm Tr}\, {\rm log}
\left[- \bD \bD_5 +\Lambda_i^2-i\epsilon\right]\, ,
\label{act2} \\
\A_{\rm I}&=&-i\frac{N_C}{2}
{\rm Tr}\, {\rm log}
\left[-\bD \left(\bD_5\right)^{-1}-i\epsilon\right]\, .
\label{act3}
\ee
The local term in eq (\ref{act1}) is the reminder of the
quartic quark interaction of the NJL model. After having shifted
the meson fields by an amount proportional to the current
quark masses $m_0$ it also contains the explicit breaking of
chiral symmetry. Furthermore we have retained the notion
of real and imaginary parts of the action as it would come
about in the Euclidean space formulation. This is also indicated
by the Feynman boundary conditions. When disentangling these
pieces, it is found that only the `real part' $\A_{\rm R}$ is
ultraviolet divergent. It is regularized within the Pauli--Villars
scheme according to which the conditions\footnote{In the case
of two subtractions we need at least two
cut-offs $\Lambda_1 $ and $\Lambda_2$. In the limit
$\Lambda_1  \to \Lambda_2 = \Lambda $, we have
$\sum_i c_i f(\Lambda_i^2)= f(0)-f(\Lambda^2)+\Lambda^2 f' (\Lambda^2 )$.
For instance, $\sum_i c_i \Lambda_i^{2n}=(2n-2)\Lambda^{2n}$.}
\be
c_0=1\, ,\quad \Lambda_0=0\, ,\quad \sum_{i=0}^2c_i=0
\quad {\rm and}\quad \sum_{i=0}^2c_i\Lambda_i^2=0
\label{pvcond}
\ee
hold. The `imaginary part' $\A_{\rm I}$ is conditionally
convergent, {\it i.e.} a principle value description must be
imposed for the integration over the time coordinate.
{\it A priori} this does not imply that it should not be
regularized. However, in order to correctly reproduce the
axial anomaly we are constrained to leave it unregularized.

Essentially we have added and subtracted the
(unphysical) $\bD_5$ model to the bosonized NJL model. Under
regularization the sum, ${\rm log}\,(\bD)+{\rm log}\,(\bD_5)$
is then treated differently from the difference,
${\rm log}\,(\bD)-{\rm log}\,(\bD_5)$.
In the case of the polarized nucleon structure
functions ii turns out that this special choice
of regularization nevertheless requires further specification.

\bigskip
\leftline{\large\it 2a. Vacuum and meson sectors}
\medskip

We consider two different Dirac operators in the background of
scalar ($S$) and pseudoscalar ($P$) fields \cite{Da95,RA95}
\be
i \bD &=& i\dslash - \left(S+i\gamma_5P\right)
+\vslash +\aslash\gamma_5
=:i\bDp+\vslash+\aslash\gamma_5
\label{defd} \\
i \bD_5 &=& - i\dslash - \left(S-i\gamma_5P\right)
-\vslash+\aslash\gamma_5
=:i\bDp_5-\vslash+\aslash\gamma_5\, .
\label{defd5}
\ee
Here we have also introduced external vector ($v_\mu$) and
axial--vector ($a_\mu$) fields. As noted above the functional
derivate of the action with respect to these sources will provide
the vector and axial--vector currents, respectively. For later use
we have also defined Dirac operators, $\bDp$ and $\bDp_5$, with
these fields omitted. Of course, all fields appearing in eqs
(\ref{defd}) and (\ref{defd5}) are considered to be matrix fields
in flavor space. It is worth noting that upon continuation to
Euclidean space, $\bD_5$ transforms into the hermitian
conjugate of $\bD$ \cite{RA95}.

In the vacuum sector the pseudoscalar fields vanish
while the variation of the action with respect to the
scalar field $S$ and yields the gap equation
\be
\frac{1}{2G}\left(m-m_0\right)
=-4iN_C m\sum_{i=0}^2c_i
\int\frac{d^4k}{(2\pi)^4}
\left[-k^2+m^2+\Lambda_i^2-i\epsilon\right]^{-1}\, .
\label{gap}
\ee
This equation determines the vacuum expectation value of
the scalar field $\langle S \rangle =m$ which is referred to
as the constituent quark mass. Its non--vanishing value
signals the dynamical breaking of chiral symmetry.
Next we expand the action to quadratic order
in the pion field ${\vec\pi}$. This field resides in the
non--linear representation of the meson fields on the
chiral circle
\be
{\cal M}=m\, U = m\, {\rm exp}
\left(i \frac{g}{m}\,{\vec\pi} \cdot {\vec\tau} \right)\, .
\label{chifield}
\ee
This representation also defines the chiral field $U$. The
quark--pion coupling $g$ will be specified shortly. Upon Fourier
transforming to ${\vec{\tilde\pi}}$ we find
\be
\A_{\rm NJL}=g^2\int \frac{d^4q}{(2\pi)^4}\,
{\vec{\tilde\pi}}(q) \cdot {\vec{\tilde\pi}}(-q)
\left[2N_C q^2\Pi(q^2)-\frac{1}{2G}\frac{m_0}{m}\right]
+{\cal O}\left(\vec{\pi}^4\right)\, ,
\label{A2}
\ee
with the polarization function
\be
\Pi(q^2,x)&=&-i\sum_{i=0}^2 c_i\,
\frac{d^4k}{(2\pi)^4}\,
\left[-k^2-x(1-x)q^2+m^2+\Lambda_i^2-i\epsilon\right]^{-2}
\quad {\rm and} \quad
\nonumber \\
\Pi(q^2)&=&\int_0^1 dx\, \Pi(q^2,x)\, ,
\label{specfct}
\ee
parameterizing the quark loop. The on--shell condition
for the pion relates its mass to the model
parameters
\be
m_\pi^2=\frac{1}{2G}\frac{m_0}{m}\frac{1}{2N_C\Pi(m_\pi^2)}\, .
\label{mpi}
\ee
Requiring a unit residuum at the pion pole
determines the quark--pion coupling
\be
\frac{1}{g^2}=4N_C \frac{d}{dq^2}
\left[q^2 \Pi(q^2)\right]\Bigg|_{q^2=m_\pi^2} \, .
\label{gpcoup}
\ee
The axial current is obtained from the linear coupling to
the axial--vector source $a_\mu$. Its matrix element between
the vacuum and the properly normalized one--pion state
provides the pion decay constant $f_\pi$ as a function of the
model parameters
\be
f_\pi=4N_C m g \Pi(m_\pi^2)\, .
\label{fpi}
\ee
The empirical values $m_\pi=138{\rm MeV}$ and $f_\pi=93{\rm MeV}$
are used to determine the model parameters.

\bigskip
\leftline{\large\it 2b. Soliton sector}
\medskip
In order to describe a soliton configuration we consider
static meson configurations. In that case it is suitable
to introduce a Dirac Hamiltonian $h$ via
\be
i\bDp=\beta(i\partial_t-h) \quad {\rm and}\quad
i\bDp_5=(-i\partial_t-h)\beta\, .
\label{defh}
\ee
For a given meson configuration the Hamiltonian $h$ is
diagonalized
\be
h\Psi_\alpha = \epsilon_\alpha \Psi_\alpha
\label{diagh}
\ee
yielding eigen--spinors $\Psi_\alpha$ and energy eigenvalues
$\epsilon_\alpha$. In the unit baryon number sector the well--known
hedgehog configuration minimizes the action for the meson fields.
This configuration introduces the chiral angle $\Theta(r)$ via
\be
h=\vec{\alpha}\cdot\vec{p}
+\beta\, m\, {\rm exp}
\left[i \vec{\hat{r}}\cdot\vec{\tau}\,
\gamma_5 \Theta(r)\right]\, .
\label{hedgehog}
\ee
The eigenstates $|\alpha\rangle$ of this Dirac Hamiltonian are in
particular characterized by their grand--spin quantum number \cite{Ka84}.
The grand--spin, $\vec{G}$ is the operator sum of total spin and
isospin. Since $\vec{G}$ commutes with the Dirac Hamiltonian
(\ref{hedgehog}) the state ${\rm exp}(i\pi G_2)|\alpha\rangle$ is
also an eigenstate of (\ref{hedgehog}) with energy $\epsilon_\alpha$
and grand--spin $G_\alpha$. This rotational symmetry, which actually
is a grand--spin reflection, will later be useful to simplify
matrix elements of the quark wave--functions, $\Psi_\alpha$.

For unit baryon number configurations it turns out that one distinct
level, $\Psi_{\rm val}$, is strongly bound \cite{Al96}. This level
is referred to as the valence quark state. The total energy
functional contains three pieces. The first one is due to the
explicit occupation of the valence quark level to ensure unit
baryon number. The second is the contribution of the polarized
vacuum. It is extracted from the action (\ref{act1}) by considering
an infinite time interval to discretize the eigenvalues of $\partial_t$.
The sum over these eigenvalues then becomes a spectral integral
\cite{Re89} which can be computed using Cauchy's theorem. Finally,
there is the trivial part stemming from the local part of the
action (\ref{act0}). Collecting these pieces we have \cite{Sc92,Al96}
\be
E_{\rm tot}[\Theta]&=&
\frac{N_C}{2}\left(1-{\rm sign}(\epsilon_{\rm val})\right)
\epsilon_{\rm val}
-\frac{N_C}{2}\sum_{i=0}^2 c_i \sum_\alpha
\left\{\sqrt{\epsilon_\alpha^2+\Lambda_i^2}
-\sqrt{\epsilon_\alpha^{(0)2}+\Lambda_i^2}
\right\}
\nonumber \\ && \hspace{2cm}
+m_\pi^2f_\pi^2\int d^3r \, (1-{\rm cos}(\Theta))\, .
\label{etot}
\ee
Here we have also subtracted the vacuum energy associated
with the trivial meson field configuration and made use
of the expressions obtained for $m_\pi$ and $f_\pi$ in the
preceding subsection. The soliton is then obtained as the profile
function $\Theta(r)$ which minimizes the total energy $E_{\rm tot}$
self--consistently.

At this point we have constructed a state which has unit baryon
number but neither good quantum numbers for spin and flavor. Such
states are generated by canonically quantizing the time--dependent
collective coordinates $A(t)$ which parameterize
the spin--flavor orientation of the soliton.
For a rigidly rotating soliton the Dirac operator
becomes, after transforming to the flavor rotating
frame \cite{Re89},
\be
i\bDp=A\beta\left(i\partial_t - \tauom
-h\right)A^\dagger
\quad{\rm and}\quad
i\bDp_5=A\left(-i\partial_t + \tauom
-h\right)\beta A^\dagger\, .
\label{collq1}
\ee
Actual computations involve an expansion with respect to
the angular velocities
\be
A^\dagger \frac{d}{dt} A = \frac{i}{2}\tauom\, .
\label{collq2}
\ee
According to the quantization rules, the angular velocities
are replaced by the spin operator
\be
\vec{\Omega}\longrightarrow \frac{1}{\alpha^2}\, \vec{J}\, .
\label{collq3}
\ee
The constant of proportionality is the moment of inertia
$\alpha^2$ which is calculated as a functional of the
soliton \cite{Re89}. For the present purpose we remark that
$\alpha^2$ is of the order $1/N_C$.\footnote{When formally considering
nucleon structure functions it will turn out that it is mostly
sufficient to refer to the Dirac operators as defined in
eqs (\ref{collq1}) rather than to explicitly carry out the
expansion in the angular velocities.} Hence an
expansion in $\vec{\Omega}$ is equivalent to one
in $1/N_C$. The nucleon wave--function becomes a (Wigner D)
function of the collective coordinates. A useful relation in
computing matrix elements of nucleon states is \cite{ANW}
\be
\langle N |\frac{1}{2}{\rm tr}
\left(A^\dagger\tau_i A\tau_j\right) |N\rangle =
-\frac{4}{3}\langle N | I_i J_j | N\rangle\, .
\label{collq4}
\ee

\bigskip
\stepcounter{chapter}
\leftline{\Large\it 3. Bjorken limit and scaling}
\medskip

An important consequence of the Pauli-Villars regularization is
given by the fact that in the Bjorken limit the model scales. A detailed
analysis \cite{wrg99} shows that in this limit the contribution to the Compton
scattering amplitude of the regularized Dirac determinant is given by
a second functional differentiation of
\be
\A_{\Lambda,{\rm R}}^{(2,v)}=
-i\frac{N_C}{4}\sum_{i=0}^2c_i
{\rm Tr}\,\left\{\left(-\bDp\bDp_5+\Lambda_i^2\right)^{-1}
\left[{\cal Q}^2\vslash\left(\dslash\right)^{-1}\vslash\bDp_5
-\bDp(\vslash\left(\dslash\right)^{-1}\vslash)_5
{\cal Q}^2\right]\right\}
\label{simple6}
\ee
with respect to the vector sources.
For the imaginary part of the action the expression analogous to
(\ref{simple6}) reads
\be
\A_{\Lambda,{\rm I}}^{(2,v)}=
-i\frac{N_C}{4}
{\rm Tr}\,\left\{\left(-\bDp\bDp_5\right)^{-1}
\left[{\cal Q}^2\vslash\left(\dslash\right)^{-1}\vslash\bDp_5
+\bDp(\vslash\left(\dslash\right)^{-1}\vslash)_5
{\cal Q}^2\right]\right\}\, .
\label{simple7}
\ee
In both cases, it is understood that the large photon momentum runs
only through the operators in square brackets.
With $S_{\mu\rho\nu\sigma}=g_{\mu\rho}g_{\nu\sigma}
+g_{\rho\nu}g_{\mu\sigma}-g_{\mu\nu}g_{\rho\sigma}$, the
subscript `5' means
\be
\gamma_\mu\gamma_\rho\gamma_\nu
=S_{\mu\rho\nu\sigma}\gamma^\sigma
-i\epsilon_{\mu\rho\nu\sigma}\gamma^\sigma\gamma^5
\quad {\rm while} \quad
(\gamma_\mu\gamma_\rho\gamma_\nu)_5
=S_{\mu\rho\nu\sigma}\gamma^\sigma+
i\epsilon_{\mu\rho\nu\sigma}\gamma^\sigma\gamma^5\, .
\label{defsign}
\ee
The fact that the sum rules enforce this extension of the
regularization scheme is not all surprising since
the derivative operator $i\dslash$ fixes the Noether
currents. Rather it is imposed and a consequence of the `sum rules'
of the model defined by $\bD_5$. The $\bD_5$ model, which is not
physical, has been introduced as a device to allow for a
regularization which maintains the anomaly structure of the
underlying theory. Hence, further specification of this
regularization prescription is demanded in order to formulate a fully
consistent model. It should be stressed that this issue is
not specific to the Pauli--Villars scheme but rather all schemes
which do regularize the sum, ${\rm log}\,(\bD)+{\rm log}\,(\bD_5)$
but not the difference, ${\rm log}\,(\bD)-{\rm log}\,(\bD_5)$
will require the specification (\ref{defsign}). Since only the
polarized, {\it i.e.} spin dependent, structure functions are effected,
this issue has not shown up when computing the pion structure functions
in the Pauli--Villars scheme.
In the unregularized case ($\Lambda_i\equiv0$)
the contributions associated to the expansion of $\bD_5$ cancel in the sum
(\ref{simple6}) and (\ref{simple7}) leaving
\be
\A^{(2,v)}=
i\frac{N_C}{2}
{\rm Tr}\,\left\{\left(\bDp\right)^{-1}
\left[{\cal Q}^2\vslash\left(\dslash\right)^{-1}
\vslash\right]\right\}\, .
\label{simple8}
\ee
Finally, it should be mentioned that
the property of scaling is not an automatic  consequence of any
regularization. For
instance, in the Proper-Time regularization, there appear logarithmic
corrections since the imaginary part of the one loop diagrams is not
regularized.

\bigskip
\stepcounter{chapter}
\leftline{\Large\it 4. The regularized structure functions}
\medskip

\bigskip
\leftline{\large\it 4a. Pion structure functions}
\medskip

Working through the expressions, one obtains after some calculation an
explicit expression for the hadronic tensor. For the pion we have
\be
 f_1 (x) = \Big( {5\over 9 } \Big)
 4N_C g^2 \frac{d}{dq^2}
\left[q^2 \Pi(q^2,x)\right]\Bigg|_{q^2=m_\pi^2} \, .
\label{pionst}
\ee
which is trivially normalized. In the chiral limit $m_\pi \to 0$ we get
\be
 f_1 (x) = \Big( {5\over 9} \Big) \theta(x) \theta(1-x)
\ee
This result is surprisingly simple for it does not seem to depend
strongly on any dynamical assumptions, and it is equivalent to set
the parton distributions, $ u_\pi (x) = \bar d_\pi (x) = 1 $ for
$\pi^+ $. In addition, if DGLAP evolution
at LO \cite{Da95} and NLO \cite{RA99} is undertaken a very satisfactory
description of valence quark distributions in the pion as extracted in
ref.\cite{SM92} is obtained. One should say however, that the gluon and sea
quark distributions are too steep in the low x region and too flat in the
high x region \cite{RA99}. In ref.\cite{Da95}, parton distributions for 
the kaon are also computed.

\bigskip
\leftline{\large\it 4b. Nucleon structure functions}
\medskip

Here we will only discuss the contribution of the
polarized vacuum to the nucleon structure functions
on a formal and conceptual level, since up to now the final formulas 
have not been numerically evaluated. 
The contribution of the distinct valence level, which
is not effected by the regularization, {\it cf.}
eq (\ref{etot}), has previously been detailed \cite{Ja75,We96}
and numerically evaluated.  
After some calculation, we obtain for the nucleon \cite{wrg99}
\be
W_{\mu\nu}(q)&=&-iM_N\frac{N_C}{4}
\int \frac{d\omega}{2\pi}\sum_\alpha \int d^3 \xi
\int \frac{d\lambda}{2\pi}\, {\rm e}^{iM_Nx\lambda}
\label{wten} \hspace{6cm}~
\\* && \hspace{2.0cm}\times\Big\langle N\Big|
\Bigg\{\Big[\bar{\Psi}_\alpha({\vec\xi}){\cal Q}_A^2
\gamma_\mu\nslash\gamma_\nu\Psi_\alpha(\xipl)
{\rm e}^{-i\lambda\omega}
\nonumber \\ && \hspace{4cm}
-\bar{\Psi}_\alpha({\vec\xi}){\cal Q}_A^2
\gamma_\nu\nslash\gamma_\mu
\Psi_\alpha(\ximl)
{\rm e}^{i\lambda\omega}\Big]
f_\alpha^+(\omega)\Big|_{\rm pole}
\nonumber \\ && \hspace{3.0cm}
+\Big[\bar{\Psi}_\alpha({\vec\xi}){\cal Q}_A^2
(\gamma_\mu\nslash\gamma_\nu)_5
\Psi_\alpha(\ximl)
{\rm e}^{-i\lambda\omega}
\nonumber \\ && \hspace{4cm}
-\bar{\Psi}_\alpha({\vec\xi}){\cal Q}_A^2
(\gamma_\nu\nslash\gamma_\mu)_5
\Psi_\alpha(\xipl)
{\rm e}^{i\lambda\omega}\Big]f_\alpha^-(\omega)\Big|_{\rm pole}
\Bigg\}\Big| N\Big\rangle\, ,
\nonumber
\label{hadten}
\ee
with the light-cone vector $n=(1,0,0,1)$ and the spectral functions
\be
f_\alpha^\pm(\omega)=\sum_{i=0}^2 c_i \frac{\omega\pm\epsilon_\alpha}
{\omega^2-\epsilon_\alpha^2-\Lambda_i^2+i\epsilon}
\pm\frac{\omega\pm\epsilon_\alpha}
{\omega^2-\epsilon_\alpha^2+i\epsilon}\, ,
\label{sfct}
\ee
and "pole" means computing the residue contribution of the poles of
the spectral functions. The previous equations are similar to the
decomposition into quark and
anti--quark distributions. In the former expression Eq.(\ref{hadten}), 
the sum is over the continuum Dirac spectrum. 
In the unregularized case,
$f_\alpha^-(\omega)=0$ while $f_\alpha^+(\omega)\big|_{\rm pole}=-
4\pi i\delta(\omega-\epsilon_\alpha)$. Apparently the
hadronic tensor then indeed becomes a sum of quark and
anti--quark distributions. However, in the Pauli--Villars
regularized scheme, we have additional contributions from
quark and anti--quark distributions with dispersion relations
which also contain the cut--offs, $\Lambda_i$. Hence they
differ from those dispersion relations na\"{\i}vely expected from
the solutions of the Dirac equation (\ref{diagh}). From this expression
structure functions can be directly obtained by contracting with
appropriate projectors.

Note that within the Bjorken limit the Callan--Gross
relation, $f_2(x)=2xf_1(x)$, is automatically fulfilled.
The unpolarized structure function $f_1(x)$ then becomes
\be
f_1(x)&=&-iM_N\frac{N_C}{2}\int \frac{d\omega}{2\pi}
\sum_\alpha \int d^3\xi \int \frac{d\lambda}{2\pi}\,
{\rm e}^{iM_Nx\lambda}
\left(\sum_{i=0}^2c_i\frac{\omega+\epsilon_\alpha}
{\omega^2-\epsilon_\alpha^2-\Lambda_i^2+i\epsilon}\right)_{\rm pole}
\nonumber\hspace{2cm}~ \\ &&\hspace{0.3cm}\times
\Big\langle N\Big|
\bar{\Psi}_\alpha(\vec{\xi}){\cal Q}^2_A\nslash
\Psi_\alpha(\xipl){\rm e}^{-i\omega\lambda}
-\bar{\Psi}_\alpha(\vec{\xi}){\cal Q}^2_A\nslash
\Psi_\alpha(\ximl)){\rm e}^{i\omega\lambda}
\Big| N\Big\rangle
\nonumber \\
&=&i\frac{5}{36}M_NN_C\int \frac{d\omega}{2\pi}
\sum_\alpha \int \frac{d\lambda}{2\pi}\,
{\rm e}^{iM_Nx\lambda}
\left(\sum_{i=0}^2c_i\frac{\omega+\epsilon_\alpha}
{\omega^2-\epsilon_\alpha^2-\Lambda_i^2+i\epsilon}\right)_{\rm pole}
\label{f1x} \\* &&\hspace{0.3cm}\times
\int d^3\xi\,\left\{ \Psi^\dagger_\alpha(\vec{\xi})
\left(1-\alpha_3\right)\Psi_\alpha(\xipl)
{\rm e}^{-i\omega\lambda}
-\Psi^\dagger_\alpha(\vec{\xi})
\left(1-\alpha_3\right)\Psi_\alpha(\ximl)
{\rm e}^{i\omega\lambda}\right\}
\nonumber
\ee
to leading order in $1/N_C$. The structure
function which enters the Gottfried sum rule \cite{Go67} should not be
regularized in contrast to previous studies. In ref \cite{Wa98}
this structure function has been treated analogously to the one
of neutrino nucleon scattering associated with the Adler sum rule.
As discussed in ref.\cite{wrg99} the latter indeed undergoes
regularization.
This example clearly exhibits that obtaining the formal expressions
for structure functions from the defining action is unavoidable
in cases when there is no relation to a static nucleon property.
Similar expressions to leading order in $1/N_C$ for the polarized
structure functions, $g_1(x)$ and $g_2(x)$ can also be found in ref.
\cite{wrg99}. Since the soliton breaks translational invariance, these 
structure functions extend out of the interval $ 0 < x < 1 $. 
The issue of restoring the proper support has been addressed in 
ref.\cite{Ga98} by going to the infinite momentum frame.

Sum rules relate moments of the structure functions to static
properties of hadrons and a consistently formulated model is
required to satisfy these sum rules. Static properties
are obtained by computing matrix elements of the symmetry
currents. In ref.\cite{wrg99}, it has been shown that 
the prescription (\ref{defsign}) complies with the required sum rules 
for polarized and unpolarized structure functions. 

\bigskip
\stepcounter{chapter}
\leftline{\Large\it 5. Phase-Space considerations}
\medskip

The result, that the pion distribution function becomes unity in the
chiral limit, seems to be a rather remarkable one, because it looks
independent of any detailed dynamics. In addition, if a DGLAP
evolution for the valence part is undertaken to leading order
\cite{Da95} and NLO \cite{RA99} the agreement with experimental
parameterizations is striking.
The interesting aspect here is that within our calculation we are
treating the pion as a composite $q\bar q$  Goldstone boson. We will
derive the same result by a rather crude method. The probability $F(x)$
(normalized to unity) that a quark in a hadron with N constituents has a
momentum fraction between $x$ and $x+dx $ is given by
\be
F(x)= \int dx_1 \dots dx_N \delta (\sum_{i=1}^N x_i  -1 ) f(x_1, \dots,
x_N ) {1\over N} \sum_{i=1}^N  \delta(x-x_i)
\ee
where  $ f(x_1 , \dots, x_N ) $ is the probability distribution of the
N quarks in the hadron. If we make the choice  $ f(x_1 , \dots, x_N ) =1 $, 
then we find
\be
F(x)= (N-1)(1-x)^{N-2} \, . 
\ee
We see that in the case $N=2$, we have $F_\pi (x)=1$, which was the
result we obtained previously for the pion. It is of course very
tempting to look at the
nucleon case ($N=3$), for which we get $ F_N (x)= 2 (1-x) $. After
DGLAP evolution we get a reasonably good description of the valence
part $ V(x)=[ u(x)+d(x)-\bar u (x) -\bar d(x) ]/3 $ (see figure), much better 
than what it is obtained in much more sophisticated
calculations \cite{model}. 
The agreement
is not surprising if one looks at the low energy parameterizations of
GRV 95' in ref. \cite{Gl95}, where one sees that as $x \to    1 $
the valence contribution behaves as $\sim (1-x) $. Actually, after
evolution one finds that for $Q^2 = 4 {\rm GeV}^2 $ the distribution
function behaves as $ F(x) \sim (1-x)^{N-2 + 1.2} $ which only agrees 
very roughly with the counting rules found long ago \cite{bf73}, $F(x) \sim
(1-x)^{2M-1}$, with $M=2$ for the nucleon. It is fair to say that the 
scale at which these counting rules were derived was never made clear, 
and that structure functions are not experimentally known for $x > 0.75$.
It is also fair to say that the
analysis above was already envisaged in the original paper of Bjorken
and Paschos \cite{bp69} although to that time perturbative QCD
evolution equations were not known, so they tried to build the full
$Q^2 $ dependent structure function by making a judicious choice of
$f (x_1 , \dots , x_N ) $. Although, the results found in this section
should be analyzed with care, it seems that the bulk of the valence
distributions is understood in terms of the number of constituents only.
Nevertheless, in both cases the gluon and sea distributions are too
steep at low x and too flat at high x \cite{RA99}. In addition, 
the $f_2$ is reasonably well described at LO and NLO at $Q^2 = 5 
{\rm GeV}`2 $ between $ 0.3 < x < 0.75 $ \cite{RA99}. 

\begin{figure}
\centerline{
\epsfig{figure=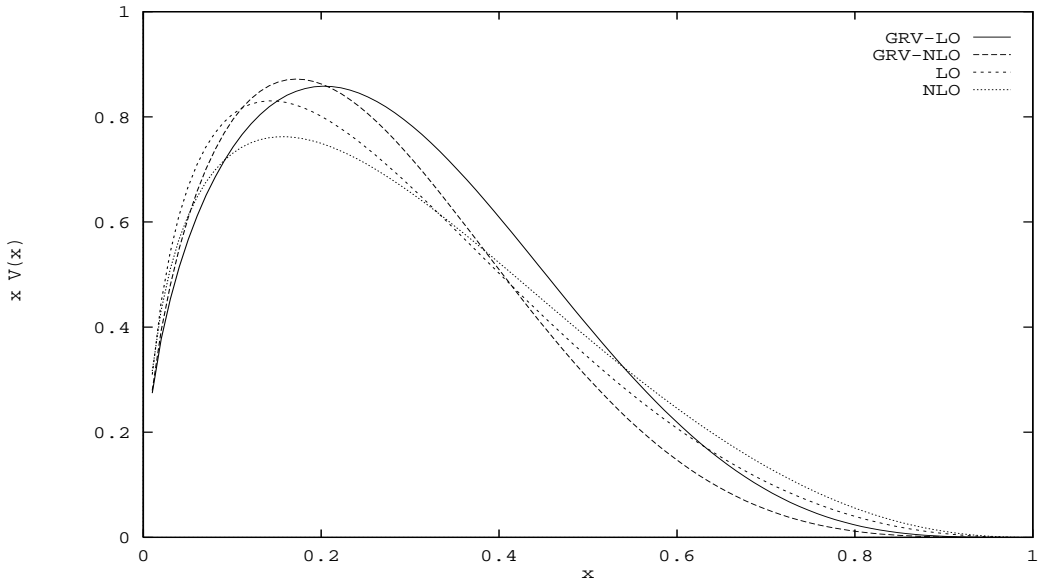,height=8cm,width=10cm}}
\caption{ \sf Valence quark distribution function in the Nucleon 
$ x V(x)=x [ u(x)+d(x)-\bar u (x) -\bar d(x) ]/3 $  computed at
LO and NLO at $Q^2 = 4 {\rm GeV}^2 $. The labels LO and NLO correspond to 
a initial distribution function $ x V(x) = 2 x (1-x) $ obtained by simple 
phase space considerations and DGLAP evolved to the scale 
$Q^2 = 4 {\rm GeV}^2 $ (see main text). The labels GRV-LO and GRV-NLO are 
the ``experimental'' analysis of ref.~\protect\cite{Gl95} at the same scale.}
\end{figure}

\newpage
\bigskip
\stepcounter{chapter}
\leftline{\Large\it 6. Can low energy models be matched to perturbative
QCD ? }
\medskip

To finish the discussion let us analyze whether or not low energy
models can be  matched to perturbative QCD in a reasonable way.
The traditional recipe for this kind of calculations is that after
the low energy model structure functions are computed, a subsequent
DGLAP evolution is undertaken. Since these models are thought
not to provide a gluon or sea distribution, the initial condition
consists of taking only the valence part, the radiative corrections
due to gluons and sea quarks, are automatically generated by the
solution of DGLAP equations. I have looked at this problem from a
different point of view \cite{RA98}. Namely, I have taken the parton
distribution functions at very high energies, $Q^2 = m_b^2 $,
and I  have evolved them downwards in energy, until some structure
function becomes negative, in the interval $ 0.01 < x < 1 $. This seems
a natural place to stop evolution, since going below would violate
unitarity. The relevant question to ask is whether or not around
this point the gluon and the sea distributions are neglegeable. There
is a subtlety in the procedure, since NLO DGLAP equations are often
used in a form which is not manifestly ``reversible'' , i.e. the
property $ U(Q_1^2 , Q_2^2 ) U(Q_2^2 , Q_3^2 ) = U(Q_1^2 , Q_3^2 ) $ is
not fulfilled. After properly handling this point I have found that
the unitarity limit takes place at the scale $Q_0 = 580 {\rm MeV} $. At
this scale one gets
\be
\langle x \rangle_{\rm val} = 0.45 \, , \quad
\langle x \rangle_{\rm gluons} = 0.36 \, , \quad
\langle x \rangle_{\rm sea} = 0.19 \, , \quad \alpha_S = 0.48
\ee
From the point of view of perturbation theory this scale is
certainly more reliable than that required by the valence quark model
point (defined as $ \langle x \rangle_{\rm val} = 1 $,
$ \langle x \rangle_{\rm gluons} + \langle x \rangle_{\rm sea} = 0 $
and which requires $\alpha_S = 1.80$ ), and it is still having a
non-negligible gluon and antiquark distribution. Thus, there seems to
be a gap between the low energy nucleon model and the perturbative QCD
result. In perturbative  QCD there exists  a systematic expansion in
powers of $\alpha_S $. In the quark model such a systematic expansion
seems to be missing, and therefore "improvements"
of it are a bit arbitrary. What is needed is a better
understanding on how to systematically improve the low energy model
including gluons and antiquarks to be really able to work on the low
energy side and hopefully fill in this gap.

\bigskip
\leftline{\Large\it Acknowledgments}
\medskip
It is a pleasure to thank the organizers of the Workshop, S. Circa, B.
Golli and M. Rosina for the stimulating atmosphere provided at Bled.
I also thank  H. Weigel, L. Gamberg and R. M. Davidson for the fruitful
collaboration. This work is supported by the Spanish DGES grant
no. PB95-1204 and the Junta de Andaluc{\'{\i}}a grant no. FQM0225.

\end{document}